\documentclass[prl,twocolumn,showpacs,preprintnumbers,amsmath,amssymb,floatfix]{revtex4}

\usepackage{graphicx}
\usepackage{dcolumn}
\usepackage{bm}

\begin{document}

\preprint{CLNS 02/1804}

\title{Revised $\alpha^4$ term of lepton $g-2$ from the Feynman diagrams

containing an internal light-by-light scattering subdiagram}

\author{T. Kinoshita}
\email{tk@hepth.cornell.edu}
\affiliation{ Laboratory for Elementary Particle Physics,
Cornell University, Ithaca, NY  U.S.A.  14853 }

\author{M. Nio}
\email{nio@riken.go.jp}
\affiliation{Theoretical Physics Laboratory,
RIKEN, Wako, Saitama, Japan 351-0198 }


\date{\today}

\begin{abstract}
The $\alpha^4$ contribution to the lepton $g-2$
from a gauge-invariant set of 18 Feynman diagrams containing a
light-by-light
scattering subdiagram internally has been reevaluated by a method
independent of the previous approach.
Comparison of two methods revealed a program error in the first
version.  Correcting this error, the contributions of these 18 diagrams
become -0.990~72~(10)$(\alpha/\pi)^4$ and -4.432~43~(58)$(\alpha/\pi)^4$
for the electron and
muon $g-2$, respectively.
The correction is not large enough to affect the comparison between
theory and experiment for the muon $g-2$, but it does alter the
inferred value for the fine structure constant $\alpha^{-1}$ by 6
ppb.

\end{abstract}


\maketitle

Precise theoretical and experimental values
of lepton anomalous magnetic moments $(a_l=(g_l -2)/2$)
provide one of the most stringent tests of QED \cite{HK}.
%
%
%
%
In case of the electron (positron), the experimental value
did reach the precision of 4.3  ppb \cite{Dehmelt}.
Currently, the theoretical  uncertainty  is dominated by
that of the fine structure constant $\alpha$. The most precise $\alpha$
available now is  from the atom interferometry experiment \cite{Chu},
which has 7.4 ppb precision.
Since the muon anomalous magnetic moment
$a_\mu$ is sensitive to short-distance physics,
high precision measurement (1.3 ppm) of
$a_\mu$ at Brookhaven National Laboratory
may be able to  open the first window to "new physics" \cite{BNLexpt}.
Before taking the discovery of $``$new physics" in the muon $g-2$
seriously, however, we must make sure that the old physics, namely the
Standard Model, is known with sufficiently high precision.

The largest source of theoretical uncertainty (0.7 ppm)
for $a_\mu$ is the hadronic contribution \cite{hadronvp, hadronlbyl}.
Unfortunately we are currently unable to deal with the
hadronic correction from first principles
because of nonperturbative nature of QCD.
On the other hand, the QED correction
can be treated precisely by perturbation theory.
In order to achieve the precision comparable to that of measurements,
the QED calculation for lepton $g-2$
must include terms of up to eighth-order of perturbation theory.
Leading contributions of tenth-order are also relevant for $a_\mu$
\cite{kino1,karsh}.

The purpose of this paper is to correct a program error
in the previous calculation of gauge-invariant set of 18 Feynman
diagrams contributing to the $\alpha^4$ QED term \cite{kino1}.
This was accomplished by constructing alternative forms
of integrals for these diagrams.
As a consequence all 891 Feynman diagrams contributing
to the eighth-order term  of $a_e$,
and additional diagrams contributing to $a_\mu - a_e$,
have now been verified by independent calculations and/or
checked by analytic comparison with lower-order integrals.
This enables us to pursue with confidence
an order of magnitude improvement in numerical precision of
all $\alpha^4$ terms of $a_\mu$ and $a_e$.
The results will be
reported shortly elsewhere \cite{kino2,kino3}.

The contribution of the QED diagrams to $a_\mu$ can be written in the
general form

\begin{eqnarray}
a_\mu ({\rm QED}) = A_1 &+& A_2 (m_\mu/m_e) + A_2 (m_\mu /m_\tau)
\nonumber   \\
       &+& A_3 (m_\mu/m_e , m_\mu/m_\tau ) ,
\end{eqnarray}
where $m_e$, $m_\mu$, and $m_\tau$ are the masses of the
electron, muon, and tau, respectively.
A similar equation holds for $a_e$.
Throughout this article we shall use the values
$m_e$ = 0.510~998~902~(21)~$MeV/c^2$,
$m_\mu$ =105.658~357~(5)~M$eV/c^2$, and
$m_\tau$ = 1~776.99~(+29,-26)~M$eV/c^2$ \cite{hagiwara}.

The renormalizability of QED guarantees that the functions $A_1$, $A_2$,
and $A_3$ can be expanded in power series in $\alpha/\pi$ with finite
calculable coefficients:

\begin{equation}
A_i = A_i^{(2)} \left ( \frac{\alpha}{\pi} \right )
+ A_i^{(4)} \left ( \frac{\alpha}{ \pi} \right )^2
+ A_i^{(6)} \left ( \frac{\alpha}{ \pi} \right )^3
+  \ldots ,~~~i=1,2,3.
\end{equation}
$A_1^{(2)}$,
$A_1^{(4)}$ and
$A_1^{(6)}$
are known analytically \cite{remiddi}.
Most terms contributing to $A_1^{(8)}$ have not yet been
obtained by analytic means.
%
%
%
%
The current uncertainty in the value of $A_1^{(8)}$ is a consequence of
the fact that, at present, it must be obtained by numerical integration.
Its precision is being improved by an extensive computer
calculation right now \cite{kino2}.
For the purpose of evaluating $a_\mu$(QED), however, it is sufficient
to use $A_1^{(8)}$ derived from the measured value of the electron
anomaly
$a_e$ \cite{vandyck}
corrected for small contributions of muon, hadron, and weak
interactions.
The terms $A_2^{(4)} (m_\mu /m_e)$ and
$A_2^{(6)} (m_\mu /m_e)$ are known exactly \cite{laporta1}.
The situation is quite different for
$A_2^{(8)} (m_\mu /m_e)$
since most terms contributing to it are known only by numerical means.

There are altogether 469 Feynman diagrams
contributing to $A_2^{(8)} (m_\mu/m_e)$.
Of these diagrams 343 have been checked by more than one independent
methods, some of which being even analytic.
Other 108 diagrams, all of
which contain an {\it external} light-by-light scattering subdiagram,
have been checked analytically by comparison with
the exactly known sixth-order vertices \cite{kino3}.
Unfortunately, this was not the case for the remaining 18,
all generated by inserting a
light-by-light scattering subdiagram {\it internally}
in a fourth-order vertex diagram (see Fig. 1).
Here {\it external} and {\it internal} means whether
one of the attached photon lines represents an
external magnetic field or not.

\begin{figure}[t]
\includegraphics{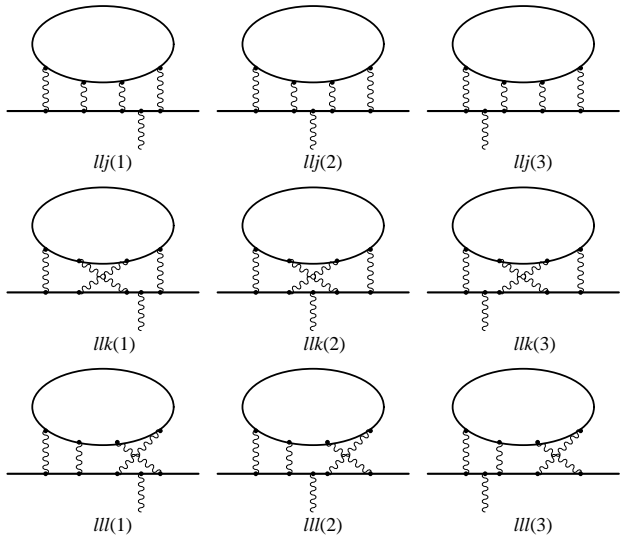}
\vspace{-0.5cm}
\caption{\label{vertexfig1}
Vertex diagrams containing a light-by-light scattering subdiagram
internally.  There are altogether 18 such diagrams.}
\end{figure}

Let us now describe how the error was discovered and corrected.
These 18 diagrams form a gauge-invariant set
and share the same basic algebraic structure.  Unfortunately,
it was not possible to examine their analytic structure
by comparison with lower-order diagrams
since they are not reducible to such diagrams
in the UV and/or IR limits.
Besides having been checked by two people working independently
\cite{oldlbyl},
the only check made was mutual consistency among these 18 diagrams.
This is not sufficient to eliminate the possibility
that they share the same program error.
Clearly, in order to enhance the credibility of
the QED calculation of $a_l$,
it is highly desirable to reevaluate these 18 diagrams
by more than one method.

\begin{figure}[t]
\includegraphics{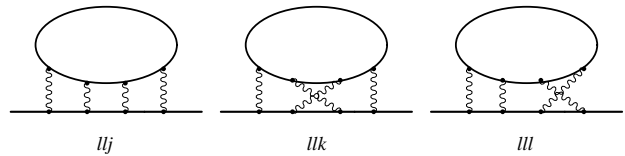}
\vspace{-0.5cm}
\caption{\label{vertexfig2}
Self-energy-like diagrams
in which lepton lines propagate in the magnetic field.
First-order terms in the weak magnetic field expansion
correspond to the diagrams of Fig. \ref{vertexfig1}.}
\end{figure}

This paper reports the consequence of our effort to reexamine
the previous result by construction of
new and independent integrals.
Before describing the new approach, let us quickly go over
the initial approach.
In that approach (which will be called {\it Version
A})\cite{oldlbyl},
we put together vertex diagrams, for instance $llj(1), llj(2), llj(3)$,
 all obtained from the self-energy-like diagram $llj$ of Fig. 2
as terms linear in the magnetic field in the weak field expansion.
This is to take advantage of their shared structure
and the tendency of partial cancellation among them.
With the help of Ward-Takahashi identity
the sum $\Lambda^\nu (p,q)$ of all vertex diagrams thus related
to a self-energy diagram $\Sigma(p)$ can be expressed as
\begin{equation}
\Lambda^{\nu}(p,q) \approx - q^{\mu} \left [ \frac{ \partial
\Lambda_{\mu}(p,q)
}{ \partial q_{\nu}} \right ]_{q=0} -\frac { \partial \Sigma (p) }{
\partial p_{\nu} },
\label{WT}
\end{equation}
where $(p+q/2)^2 = (p-q/2)^2 = m_{l}^2$.
The $g-2$ term is projected out from the right-hand side of (\ref{WT}).
In terms of Feynman parameters $z_1, z_2, \ldots z_N$,
the $n$-th order magnetic moment
derived from (\ref{WT}) has a form
\begin{eqnarray}
M^{(2n)} &=& \Biglb ( \frac{ -1 }{ 4 } \Bigrb )^n (n-1)! \int (dz)
\Biglb ( \frac{ \bm{E} + \bm{C} }{ n-1 } \frac{ 1 }{ U^2 V^{n-1} }
\nonumber  \\
&+& (\bm{N} +\bm{Z} ) \frac{ 1 }{ U^2 V^n } \Bigrb )~,
\label{WTtransform}
\end{eqnarray}
%
%
%
%
where  bold letters  $\bm{E}$, $\bm{C}$, $\bm{N}$, and $\bm{Z}$
stand for parts of the projection operator adapted to the 
first and second terms on the right-hand-side of (\ref{WT}).
$U$ is the Jacobian of transformation from momentum-space variables to
Feynman parameters.
$V^{-1}$ is obtained by combining all propagators into one
with the help of Feynman parameters.
See \cite{CK1} for precise definitions
of projection operators, $U, V$ and $(dz)$.

In order to check the validity of {\it Version A} based on Fig. 2
we evaluated these diagrams by a second method,
called {\it Version B}, which is actually a straight-forward
parametrization of individual vertex diagrams
of Fig. 1, without relying on the Ward-Takahashi identity.
In this approach it is convenient to evaluate the sum
$jkl(n)\equiv llj(n) + llk(n) +lll(n)$, where $n=1,2,3$,
since partial cancellation of singular terms occurs
resulting in less singular behavior.

Diagrams of Fig. 1 (or Fig. 2) form a (formal) gauge-invariant set.
But individual diagrams are UV-divergent,
and must be regularized in advance, for instance by dimensional
regularization,
to enforce gauge invariance.
For numerical evaluation, however, it turns out to be more convenient to
combine it with a subtractive regularization.
Let $F_m (d)$ be one of the integrals defined in $d$-dimension,
where $m$ is any one of $llj(1), \ldots, lll(3)$.
Let $G_m (d)$ be the subtraction term
containing the light-by-light scattering
tensor with zero external momenta $\Pi_{\mu \nu \sigma \rho}(0,0,0,0)$
as well as terms containing a sixth-order charge renormalization
diagram.
Let us rewrite $F_m (d)$ symbolically as
\begin{equation}
~~~~~~~~~(F_m (d)- G_m (d)) + G_m (d),
\label{rewrite}
\end{equation}
where "symbolically" means that
subtraction is performed on the integrand
before the integration is carried out.  Now,
we can  safely take the limit $d \rightarrow 4$ for the first term
since its integrand does not cause UV divergence.
Of course, the second term $G_m (d)$ is singular for $d \rightarrow 4$.
%
%
%
%
However, gauge invariance guarantees that
the sum of $G_m(d)$ over all diagrams of Fig. 1 vanishes for any
value of
dimension  $d$:
\begin{equation}
~~~~~~~~~\sum_m G_m (d) = 0.
\label{gaugesum}
\end{equation}
Thus we have only to compute $(F_m (4) - G_m (4))$ in the end.

The integrands of both {\it Version A} and {\it Version B} were
generated
by an algebraic program FORM \cite{vermasseren}.
Numerical integration is carried out by
an adaptive-iterative Monte-Carlo routine VEGAS \cite{vegas}.

When we compared the numerical results of {\it Version A} and {\it
Version B},
we were surprised to find
that their values were significantly different.
After an extensive detective work, we located
a programming error in {\it Version A},
which resulted from an incomplete implementation
of the $\bm{E}$ operation
of (\ref{WTtransform})
in the algebraic manipulation program:
It left out some terms referring
to the light-by-light loop subdiagram.
Such a program error can be readily detected
if the integral exhibits UV or IR divergence
after renormalization is carried out.
Unfortunately the particular error in the 18 diagrams
caused no divergence and escaped scrutiny of two people.
Once this error was corrected, both approaches gave
identical numerical results.
Thus we now have two sets of independent
codes for the 18 diagrams that have been fully verified.

Numerical evaluation of these diagrams
requires an enormous amount of computational effort.
A systematic algorithm of computation to minimize human error
is indispensable.
Such a scheme was developed originally for the
calculation of the sixth-order lepton $g-2$ \cite{CK1,CK23},
and were thoroughly tested over the years \cite{kino4}.
It was later extended to the eighth-order
\cite{eighth}.
Early results for the eighth-order term remained rather crude for
many years.
This is mainly due to the enormous size of the integrands
which could not be handled adequately by the computers then available.
More precise values have become
available only in this decade thanks to the development of massively
parallel computer, which enabled us to vastly increase the sampling
statistics of VEGAS.

Enlarging sampling statistics, however,
amplified the difficulty caused by a previously poorly understood
problem in estimating errors in a computer calculation.
This arises from the fact that computer calculation always deals
with a finite number of digits.
This means that the error estimate based on
the assumption of normal distribution of errors must be
modified to take the effect of rounding-off of digits into account.
In our formulation in which subtractive renormalization of UV
divergence as well as separation of IR
divergences are carried out on the computer, this
{\it digit deficiency problem} can distort error
estimates seriously, or even prevent further iteration.
Besides increasing the number of effective digits
from real*8 to real*16 arithmetic,
which was the most obvious and effective cure, various methods
had to be devised to deal with the {\it digit deficiency error}
\cite{nio3}.

\begin{table}
\caption{\label{table1} Muon  $g-2$ contributions from the diagrams
of Fig. 1.
In $ Version$ $A$ the Ward-Takahashi-summed
$llj \equiv llj(1)+llj(2)+llj(3)$,
$llk \equiv llk(1)+llk(2)+llk(3)$,
 and $lll \equiv lll(1)+lll(2)+lll(3)$ are
calculated, while in $Version$ $B$
$jkl(1,3) \equiv llj(1)+llj(3)+llk(1)+llk(3)+lll(1)+lll(3)$,
and  $jkl(2) \equiv llj(2)+llk(2)+lll(2)$ are calculated.
}
\begin{ruledtabular}
\begin{tabular}{crcr}
\multicolumn{2}{c}{ $Version$  $A$} & \multicolumn{2}{c}{ $Version$
$B$}
\\
\hline
   $llj$& 6.389~802~(460) & $jkl$(1,3)&-3.509~978~(802)\\
   $llk$&-7.763~474~(537) &  $jkl$(2)&-0.921~589~(~89)~\\
   $lll$&-3.059~704~(452) &      &         \\
\hline
sum        &---4.433~376~(840) & sum     &-4.431~567~(806)\\
\end{tabular}
\end{ruledtabular}
\end{table}

\begin{table}
\caption{\label{table2} Electron $g-2$ contributions from the diagrams
of
Fig. 1.
In $ Version$ $A$ the Ward-Takahashi-summed
$llj \equiv llj(1)+llj(2)+llj(3)$,
$llk \equiv llk(1)+llk(2)+llk(3)$, and
$lll \equiv lll(1)+lll(2)+lll(3)$ are
calculated, while in $Version$ $B$
$jkl(1,3) \equiv llj(1)+llj(3)+llk(1)+llk(3)+lll(1)+lll(3)$,
and  $jkl(2) \equiv llj(2)+llk(2)+lll(2)$ are calculated.
}
\begin{ruledtabular}
\begin{tabular}{crcr}
\multicolumn{2}{c}{ $Version$  $A$} & \multicolumn{2}{c}{ $Version$
$B$}
\\
\hline
 $llj$& 2.551~223~(78) & $jkl$(1,3)&-0.872~717~(138)\\
 $llk$&-1.873~801~(72) & $jkl$(2) &-0.117~959~(~28)\\
 $lll$&-1.668~182~(80) &          &     \\
\hline
sum&-0.990~760~(133) & sum  & -0.990~675~(141) \\
\end{tabular}
\end{ruledtabular}
\end{table}
New results of numerical integration of $a_\mu$
by {\it Versions} $A$ and $B$ are listed in Table I.
The values of $llj, llk,$ and $lll$ listed
were obtained using $5 \times 10^9$ sampling points
per iteration and iterated 110, 219, 220 times, respectively.
They were evaluated on $v1$ cluster at Cornell Theory Center.
The calculation of {\it Version B} was carried out on Fujitsu VPP700E
at the  Computer and Information Division  of RIKEN.
For $jkl$(1,3)
$4.6\times 10^9 $ sampling points per iteration were used for
131 iterations.  The program $jkl$(2) shows less singular behavior.
It was evaluated using $4.6\times 10^9 $ sampling points per iteration
and iterated 60 times.

Results for the electron are listed in Table II.
The values of $llj, llk$, and $lll$ listed
were obtained using $2 \times 10^9$ sampling points
per iteration and iterated 160, 220, 180 times, respectively,
on $v1$ cluster at Cornell Theory Center.
For $jkl$(1,3) $4.6 \times 10^9$ sampling points per iteration were
used for 60 iterations on VPP700E. For $jkl$(2)
$4.6 \times 10^9 $ sampling points per iteration were used
and   iterated 60 times.

Combining the results from Tables I and II,
which we treat as statistically independent,
we obtain the best estimate of the contribution
to $a_\mu$ from the 18 Feynman diagrams of Fig. 1:
\begin{equation}
 a_{IV(d)}^{(8)\mu}=-4.432~43~(58) \left (\frac{\alpha }{ \pi} \right
)^4
,
\label{a_mu}
\end{equation}
and a corresponding result for $a_e$
\begin{equation}
 a_{IV(d)}^{(8)e}=-0.990~72~(10) \left (\frac{\alpha }{ \pi} \right )^4
.
\label{a_e}
\end{equation}
Here, superscripts $(8)\mu$ and $(8)e$
refer to the eight-order muon and electron $g-2$,
the subscript $IV$ to the group of all diagrams containing
light-by-light scattering subdiagrams, and
$(d)$ to its subgroup shown in Fig. 1,
which consists of all diagrams containing an internal light-by-light
scattering diagram.

The new results (\ref{a_mu}) and (\ref{a_e}) supersede the earlier
values $-$3.4387 (533) $(\alpha/\pi )^4$ \cite{kino1}
and $-$0.7503 (60) $(\alpha/\pi )^4$ \cite{kino5}, respectively.
The effect of this modification on $a_\mu$(QED)
is less than $1 \%$ of the overall eighth-order term
(which is of the order of 130 $(\alpha /\pi )^4$),
and thus does not affect comparison of experiment and theory
significantly.
On the other hand, the effect on $a_e$ is $ \sim -7.0 \times 10^{-12}$
which is about 16 \% of the entire eighth-order term
and is larger than the measurement uncertainty $4.3 \times 10^{-12}$.
As a consequence, it reduces
the inverse fine structure constant $\alpha^{-1}$
obtained from theory and measurement of $a_e$ by $ \sim 0.82 \times
10^{-6}$ or $ \sim 6$ ppb.
Currently all $\alpha^4$ terms are being upgraded
by an extensive numerical integration.
Precise values of $a_\mu$(QED) and $a_e$(QED)
including these terms will be reported in
\cite{kino2,kino3}.

\begin{acknowledgments}
T. K. should like to thank RIKEN for the hospitality extended to him
while a part of this work was carried out.
The part of the material
presented by T. K. is based upon work supported by the National
Science Foundation under Grant No. PHY-0098631.
The numerical work for {\it Version A} has been carried out on
the $``$velocity cluster" computer at Cornell Theory Center,
which receives funding from Cornell University, New York State,
the National Center for Research Resources at the National Institute of
Health,
the National Science Foundation, the Defense Department
Modernization Program, the United States Department of Agriculture,
and the corporate partners.
The numerical work of {\it Version B} has been carried out
on the Fujitsu VPP700E  at Computer and Information Division of RIKEN.
\end{acknowledgments}

\end{document}